\documentclass[twocolumn,showpacs,preprintnumbers,amsmath,amssymb,superscriptaddress]{revtex4}

\usepackage{graphicx}
\usepackage{dcolumn}
\usepackage{bm}

\begin{document}

\preprint{APS/123-QED}

\title{Magnetic field-induced quantum superconductor-insulator transition in $Nb_{0.15}Si_{0.85}$}

\author{H. Aubin}
\email{Herve.Aubin@espci.fr}
\affiliation{Laboratoire de Physique
Quantique, ESPCI (UPR5-CNRS) 10 Rue Vauquelin, 75231 Paris,
France}
\author{C.A. Marrache-Kikuchi}
\affiliation{ CSNSM, IN2P3-CNRS, B\^{a}timent 108, 91405 Orsay
Campus, France }
\author{A. Pourret}
\author{K. Behnia}
\affiliation{Laboratoire de Physique Quantique, ESPCI (UPR5-CNRS)
10 Rue Vauquelin, 75231 Paris, France}
\author{L. Berg\'{e}}
\author{L. Dumoulin}
\affiliation{ CSNSM, IN2P3-CNRS, B\^{a}timent 108, 91405 Orsay
Campus, France }
\author{J. Lesueur}
\affiliation{Laboratoire de Physique Quantique, ESPCI (UPR5-CNRS)
10 Rue Vauquelin, 75231 Paris, France}

\date{\today}

\begin{abstract}
A study of magnetic-field tuned superconductor-insulator
transitions in amorphous $Nb_{0.15}Si_{0.85}$ thin films shows
that quantum superconductor-insulator transitions are
characterized by an unambiguous signature -- a kink in the
temperature profile of the critical magnetic field. Using this
criterion, we show that the nature of the magnetic-field tuned
superconductor-insulator transition depends on the orientation of
the field with respect to the film. For perpendicular magnetic
field, the transition is controlled by quantum fluctuations with
indications for the existence of a Bose insulator; while for
parallel magnetic field, the transition is classical, driven by
the breaking of Cooper pairs at the temperature dependent critical
field $H_{c2}$.
\end{abstract}


\maketitle

Quantum fluctuations are believed to control the critical behavior
of continuous Superconductor-Insulator Transitions (SIT) observed
in disordered thin films of various metals~\cite{Sondhi97}.
Fine-tuning of the transition can be achieved either by  applying
a perpendicular magnetic
field~\cite{Hebard90,Paalanen92,Yazdani95,Ephron96,Markovic98,Gantmakher00,Bielejec02,Sambandamurthy04}
or by varying the sheet resistance $R_{\square}$ of the films --
using film thickness\cite{Haviland89,Markovic99} or electrostatic
field\cite{Parendo05}.

These transitions have been found to be characterized by a
critical resistance -- where the Temperature Coefficient of
Resistance(TCR) $dR/dT$ changes sign -- and scaling behavior as
predicted by the so-called \textquoteleft dirty
boson\textquoteright ~model~\cite{Fisher90-prl64}. In case of
magnetic field-induced SIT, the \textquoteleft dirty
boson\textquoteright~model predicts that the SIT occurs
simultaneously with the quantum melting and condensation of the
vortex system~\cite{Fisher90-prl65,Kramer98}. Indeed, because the
dual representation of the vortex system is isomorph to a
disordered two-dimensional boson system~\cite{Nelson88}, and
because Cooper pairs in thin films can also be described by a
two-dimensional boson system, it has been shown that this model is
self-dual~\cite{Fisher90-prl65}. Thus, the superconducting state
is characterized by localized vortices in a condensate of Cooper
pairs; the insulating state, by localized Cooper pairs in a
condensate of vortices.

However, experimentally, it is difficult to distinguish this Bose
insulator from the standard fermionic insulator expected in
two-dimensional disordered fermions systems~\cite{Abrahams79}.
Moreover, because the crossing point observed in the field
dependence of the resistance curves measured at different
temperatures is generally expected whenever the sample goes from
superconducting ($dR/dT>0$) to insulating ($dR/dT<0$), the
conclusion about the quantum nature of the transition only relies
on the temperature \emph{independence} of this crossing point.
Thus, for a given experiment, it can never be definitely excluded
that the apparent temperature \emph{independence} of the crossing
point is not due to the finite resolution of measurements.

This ambiguity was particularly striking in a recent report of a
comparison of parallel and perpendicular field-tuned
SIT~\cite{Gantmakher00}. In this experiment, no features in the
data have been found that could distinguish the transitions
observed in the two magnetic field configurations; while we do
expect the nature of the transitions to be profoundly different
for the two configurations.

Indeed, as described above, in a perpendicular magnetic field, the
SIT is due to the quantum melting and condensation of the vortex
glass; however, in a parallel magnetic field, because no vortices
can be created in the sample, we expect the field-induced
transition to be classical, due to the breaking of the Cooper
pairs at $H_{c2}$.

In this letter, we report on a study of the magnetic field-tuned
SIT in $Nb_xSi_{1-x}$ amorphous thin films for two orientations of
magnetic field, parallel and perpendicular to the film plane. A
careful examination of the SIT in the two configurations clearly
shows that they hold distinct characteristics. For parallel
magnetic fields, the transition is \textquoteleft
classical\textquoteright, due to the vanishing superconducting
order parameter at the temperature dependent critical field
$H_{c2}$; whereas, for perpendicular magnetic field, the
transition is controlled by quantum fluctuations, characterized by
a temperature independent critical point $(R_c,H_c)$ and universal
scaling behavior. Furthermore, a clear kink is observed in the
temperature profile of the critical field that definitely
indicates the peculiar (quantum) nature of the transition.

Amorphous thin films of $Nb_xSi_{1-x}$ are prepared under
ultrahigh vacuum by e-beam co-evaporation of Nb and Si, with
special care over the control and homogeneity of concentrations.
Such films are known to undergo a transition from insulator to
metal with increasing Nb
concentration~\cite{Dumoulin93,Bishop85,Lee00}. For this
experiment, a series of six samples, numbered 1 to 6 in
table~\ref{tab:table1}, with stoichiometry $Nb_{0.15}Si_{0.85}$
and thicknesses ranging from $100~nm$ down to $12.5~nm$, have been
deposited onto sapphire substrates. Those films are highly stable
and no significant changes of resistance are observed after a
cycling between room and helium temperature.

Resistances were measured by employing a standard four-probe
method, with ac lock-in detection operated at $23~Hz$. The
magnitude of the current is $100~nA$, which is well within the
linear regime of the I-V characteristic. All electrical leads were
filtered at $300~K$ from RF frequency.

\begin{table}
\caption{\label{tab:table1}The samples and their parameters. $T_c$
is defined as the temperature at which the resistance is half the
normal state value. $H_c$ and $R_c$ are read from the crossing
point observed for each sample, in perpendicular magnetic field.}
\begin{ruledtabular}
\begin{tabular}{cccccccc}
$No.$ & $d$[\AA] & $R_n[\Omega]$ & $R_c[\Omega]$ & $T_{c0}[mK]$ & $H_c[kOe]$ & $\nu$ \\
\hline
1 & 500& 287 & 282 & 480 & 10.1 & 0.65\\
2 & 250& 632 & 612 & 375 & 7.7 & 0.72\\
3 & 250& 638 & 620 & 347 & 8.0 & 0.8\\
4 & 125& 1401 & 1333 & 235 & 5.5 & 0.67\\
5 & 1000& 152 & $\approx 150$ & 530 & $\approx 11.0$ & ?\\
6 & 125& 1430 & 1356 & 213 & 5.0 & 0.6\\

\end{tabular}
\end{ruledtabular}
\end{table}

\begin{figure}
\includegraphics[width=6cm]{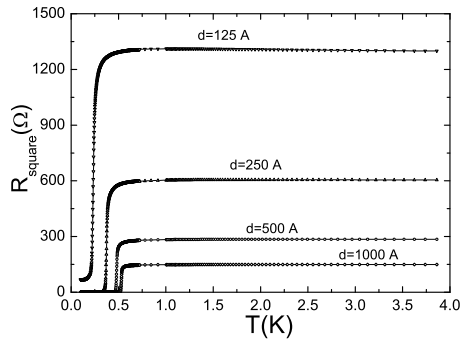}
\caption{\label{fig:fig0} $R_{\square}$ versus temperature for
four samples, numbered 1,2,4,5 in table~\ref{tab:table1}. The
superconducting transition temperature decreases with sample
thickness.}
\end{figure}

As shown in table~\ref{tab:table1}, the sheet resistance
$R_{\square}$ of the samples increases from $152~\Omega$ up to
$1430~\Omega$ when film thickness decreases from $100~nm$ down to
$12.5~nm$. It is important to note that the normal state
resistivities of these films are almost equal, which indicates
that they are homogenously disordered, with the length scale for
disorder potential only a few atomic spacings, as shown by
structural studies~\cite{KikuchiPhd}.

Figure~\ref{fig:fig0} shows the temperature dependence
$R_{\square}(T)$ for samples of different thicknesses. The
superconducting transition temperature decreases for samples of
increasing sheet resistance with no sign of reentrant behavior
characteristic of granular systems, which is another indication of
the homogeneity of the films. The decrease of the superconducting
transition temperature is most likely due to the weakening of
Coulomb screening in presence of disorder~\cite{Finkelstein87}.

\begin{figure}
\includegraphics[width=\linewidth]{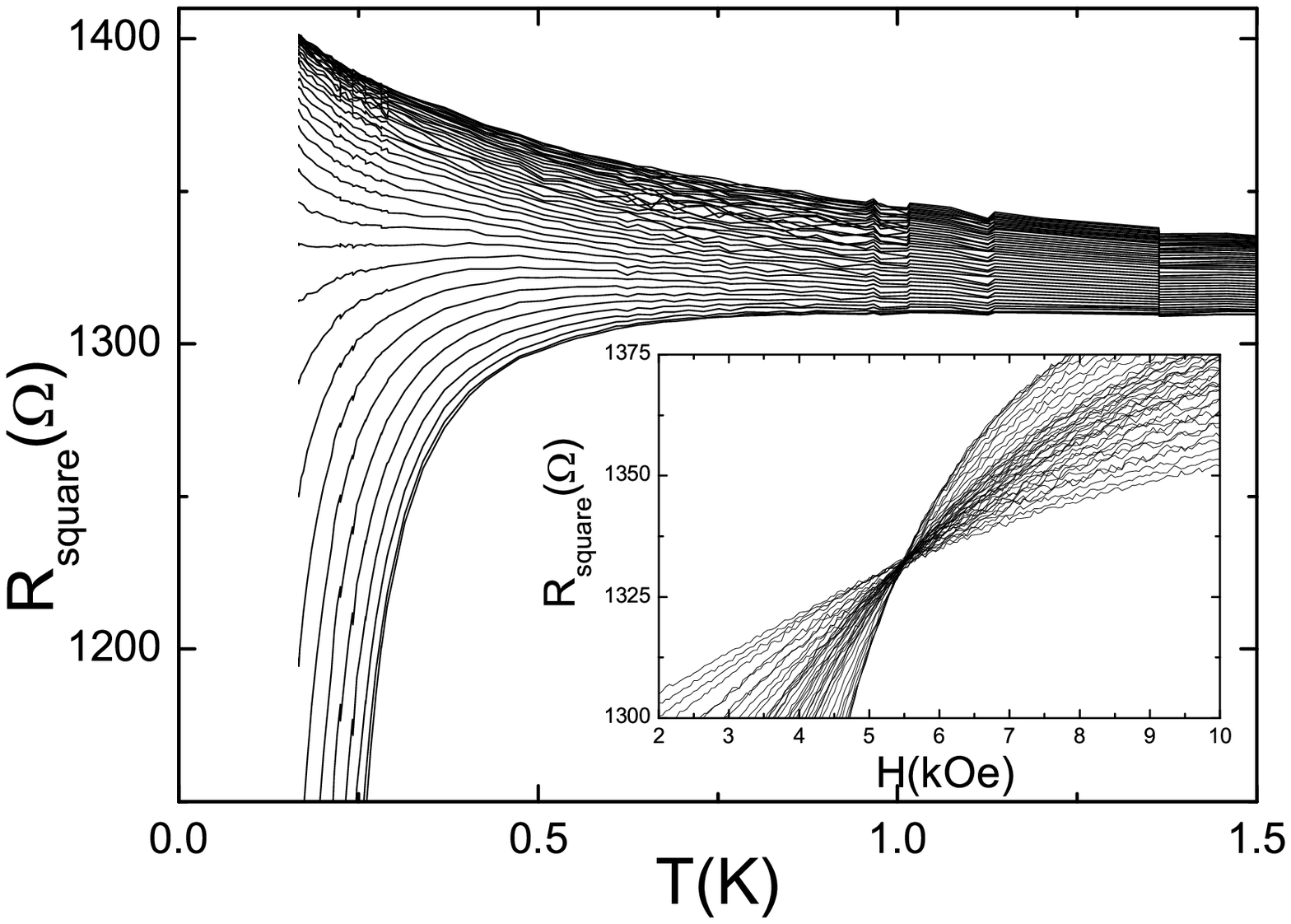}
\\
\includegraphics[width=\linewidth]{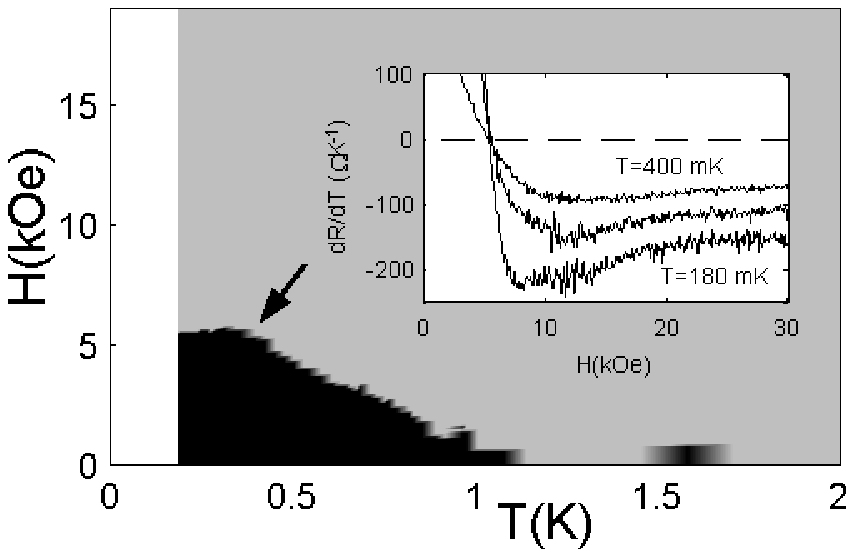}
\caption{\label{fig:fig1} Top panel: $R_{\square}$ versus
temperature of sample 4, displayed for perpendicular magnetic
field values between $0~kOe$ and $20~kOe$ by steps of $0.5~kOe$.
The TCR is positive below the critical field $H_c=5.5~kOe$ and
negative above. Top inset: $R_{\square}$ versus magnetic field for
the same sample measured at temperatures between $150~mK$ and
$400~mK$. Bottom panel: Temperature-magnetic field (T,H) color map
of TCR sign, where light (dark) grey is for negative (positive)
TCR. The border between the two regions locates the crossing point
observed in $R_{\square}$ versus H curves. The color map clearly
shows a kink about $400~mK$, as indicated by the arrow, below
which the critical field is temperature independent. Bottom inset:
TCR versus H for temperature between $180~mK$ and $400~mK$.}
\end{figure}

From now on, the data shown are measured on sample 4, $125~$\AA
thick. Figure~\ref{fig:fig1} shows the temperature dependance
$R_{\square}(T)$ for different values of a magnetic field applied
normal to the thin film. At large magnetic fields, $R_{\square}$
rises with decreasing temperature, signaling the onset of
insulating behavior in the zero temperature limit. The resistance
isotherms, measured as function of magnetic field at fixed
temperatures between $150~mK$ and $400~mK$, clearly show , inset
of figure~\ref{fig:fig1}, the existence of a critical value of the
magnetic field $H_c=5.5~kOe$ where $R_c=1333~\Omega$. As shown in
previous
reports~\cite{Hebard90,Paalanen92,Yazdani95,Ephron96,Markovic98,Gantmakher00,Bielejec02,Sambandamurthy04,Steiner05},
the insensitivity of this crossing point to temperature is the
most striking characteristic of the data, and is believed to be
the consequence of the quantum nature of the SIT. An alternative
way to display the critical point is to plot the TCR, with a
two-color scale, on the temperature-magnetic field plane, where
light (dark) gray is used for negative (positive) TCR. At the
crossing point, the TCR changes sign and is represented as the
border line between the two colored regions. The color map clearly
shows a kink in the temperature profile of the critical field,
about $400~mK$. Below this temperature, the critical field is
temperature-independent and represents, on this color-map, the
crossing point ($R_c$,$H_c$). The kink clearly defines a
temperature scale in the system -- absent from classical
transitions -- that signal the peculiar nature of the transition;
this kink may be interpreted as the temperature scale below which
quantum fluctuations of the superconducting order parameter
dominate the dynamics of the system.

\begin{figure}
\includegraphics[width=\linewidth]{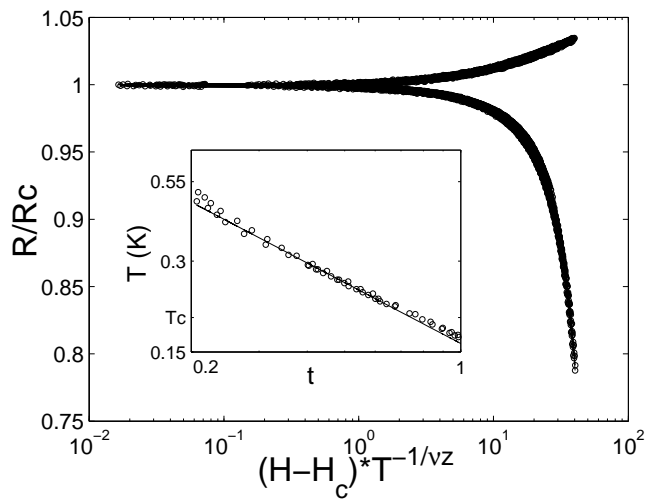}
\caption{\label{fig:fig5} Scaling of $R_{\square}/R_c$ versus
$|H-H_c|/T^{1/\nu z}$ for sample 4. All the data at temperatures
between $150~mK$ and $450~mK$, magnetic fields between $0.5*H_c$
and $1.5*H_c$ are shown here to collapse. Inset: Log-log plot of
the temperature as function of the scaling parameter $t(T)$.
Fitting of this data with $t(T)=T^{-1/\nu z}$ gives the product
$\nu z=0.67\pm0.05$.}
\end{figure}

Using the value of the critical point $(R_c,H_c)$ just determined,
we plot the ratio $R_{\square}/R_c$ against the scaling variable
$|H-H_c|/T^{1/\nu z}$, where $\nu$ is the correlation length
exponent and $z$ the dynamical-scaling exponent.

The critical exponent product $\nu z$ is obtained from the data
using a numerical minimization procedure described in
Ref.~\cite{Markovic99}. The procedure consists to plot
$R_{\square}/R_c$ versus $|H-H_c|*t(T)$ and treat $t(T)$ as an
unknown variable. The value of $t(T)$ is found when the best
collapse between data measured at temperature T and the data
measured at our lowest temperature($150~mK$) is obtained. A fit of
the data t(T) with the function $t=T^{-1/\nu z}$ gives the
exponent product $\nu z=0.67\pm0.05$; note that the value of the
exponent does not depend on the curve against which we test the
collapse of the data, and figure~\ref{fig:fig5} demonstrates the
good collapse of the data for this value of the exponent. With
increasing sample thickness, weaker insulating behavior is
observed at high field, but the main aspects of the transition --
critical point and scaling behavior -- remain up to 500\AA.
Despite the fact that the value of the critical field $H_c$ is
observed to increase and the critical resistance $R_c$ to decrease
with increasing sample thickness, the value of the exponents are
observed to remain identical, $\nu z=0.7\pm0.1$. Note that the
scaling behavior of sample 5, $1000$ \AA thick, was difficult to
analyze due its very weak insulating behavior and poorly defined
critical point.

It has been argued that, due to long-range Coulomb interactions,
the dynamical exponent should have the value $z=1$
~\cite{Fisher90-prl64}, a value confirmed by independent
measurement of $z$ and $\nu$ in MoGe~\cite{Yazdani95}. Assuming
this value $z=1$, it follows that $\nu =0.67\pm0.05$, which is
identical to exponent values obtained from the study of the
field-tuned SIT in amorphous Bi films~\cite{Markovic99} and
amorphous Be films measured with high currents~\cite{Bielejec02}.
However, exponent values $\nu\simeq1.3$ have been obtained from
field-induced SIT observed in $InO_x$~\cite{Hebard90},
MoGe~\cite{Yazdani95}, and amorphous Be films measured with low
current~\cite{Bielejec02}. The reasons behind the observed
differences are still unknown.

An exponent value $\nu\approx0.67$ agrees with numerical
simulations of the (2+1)-dimensional classical XY
model~\cite{Cha94} and the Boson-Hubbard model~\cite{Kisker97}.
This exponent is, however, inconsistent with the scaling theory of
the dirty boson model, which predicts $\nu \geq
1$~\cite{Fisher90-prl64}, and percolation-based model
~\cite{Shimshoni98}, which predicts $\nu=1.3$.

An interesting feature of the data is that compelling scaling
behavior, figure~\ref{fig:fig5}, and a well defined critical
point, figure~\ref{fig:fig1}, are observed up to $T\approx
400~mK$, above the superconducting critical temperature
$T_c=235~mK$. This indicates that the critical behavior of the SIT
is not affected by the dramatic change in the life-time of Cooper
pairs, from infinite below $T_c$ to finite values above $T_c$, in
the fluctuating regime.

The value of the critical resistances found in our experiment are
also inconsistent with the scaling theory, which predicts a
universal value $R_Q=h/4e^2\approx 6.5~k\Omega/\square$ for the
resistance at the quantum critical point. Instead, we find that
the critical resistance changes with the sample sheet resistance,
to span a large interval between $150~\Omega$ and $1333~\Omega$,
see table~\ref{tab:table1}. Such deviations of the critical
resistance from the universal value have often been
found~\cite{Yazdani95}; theoretically, it has been suggested that
models including fermionic excitations may account for the excess
conductivity at the critical point~\cite{Kapitulnik01,Galitski05},
while preserving the main characteristics of the quantum SIT : the
critical point and scaling behavior.

Despite those inconsistencies between the data and the purely
bosonic models, we will now show that the most representative
characteristics of the field-tuned quantum SIT are not found when
the same samples are submitted to a parallel magnetic field.

\begin{figure}
\includegraphics[width=\linewidth]{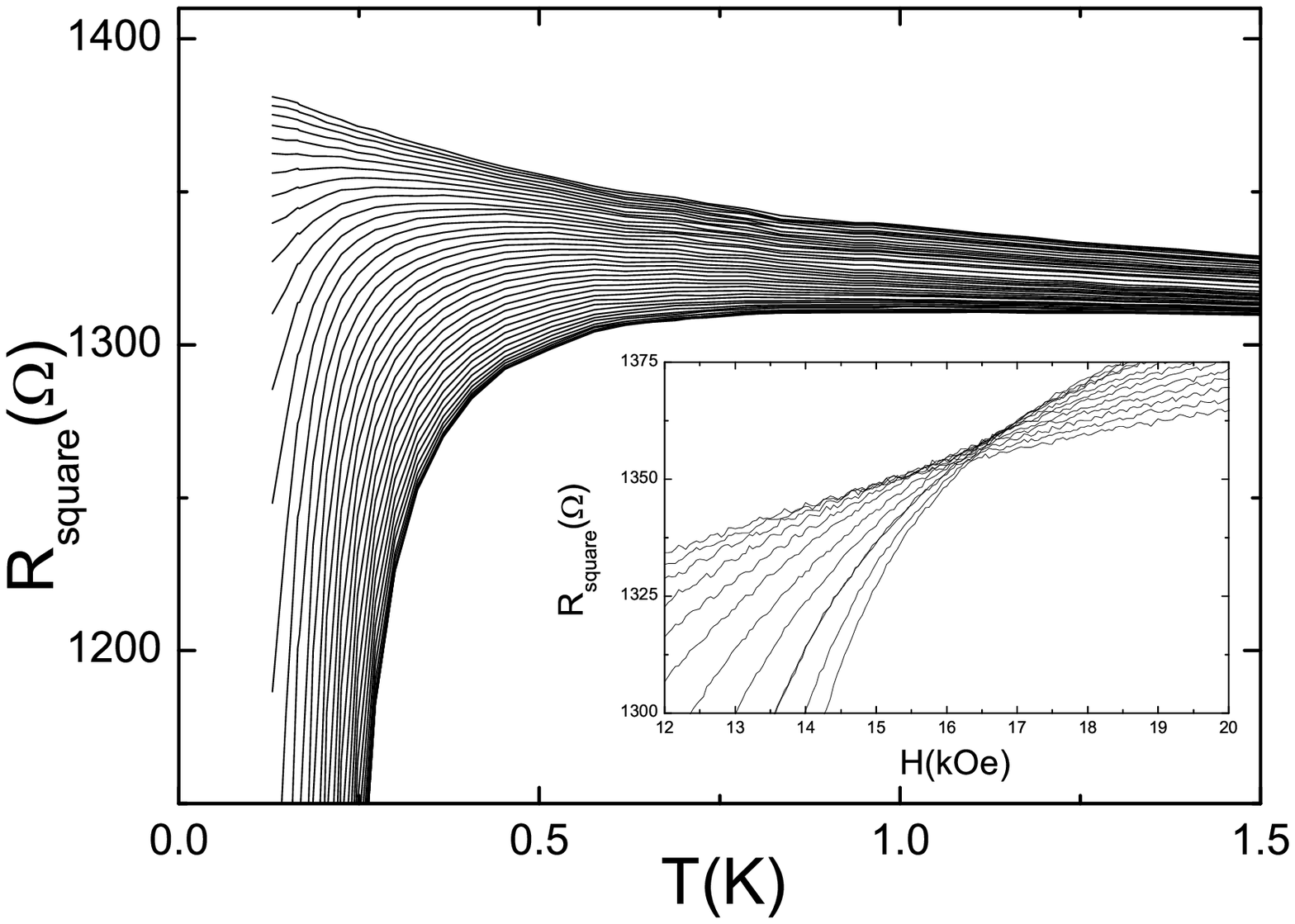}
\\
\includegraphics[width=\linewidth]{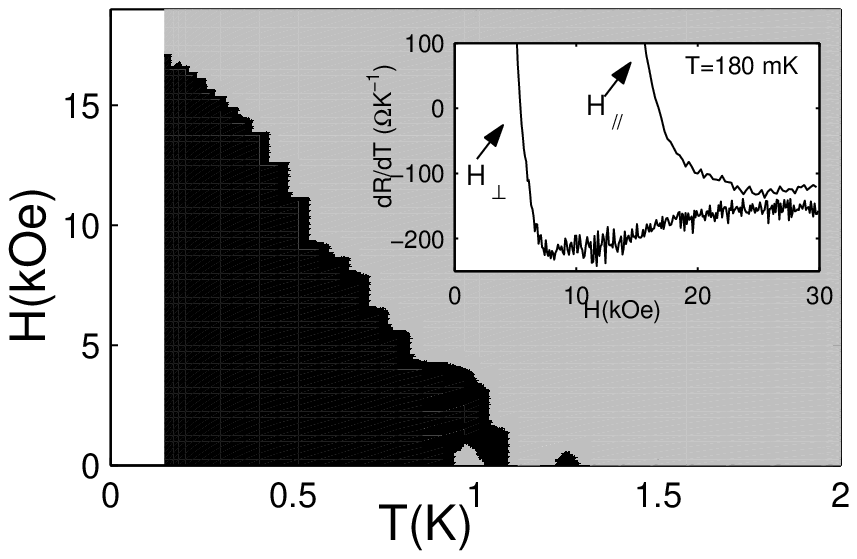}
\caption{\label{fig:fig3}Top panel: $R_{\square}$ versus
temperature of sample 4 displayed for parallel magnetic field
values between $0~kOe$ and $20~kOe$, by step of $0.5~kOe$. Top
inset: $R_{\square}$ versus magnetic field measured at
temperatures between $150~mk$ and $400~mK$. Bottom panel: (H,T)
color map of the sign of TCR, where light (dark) grey is for
negative (positive) TCR. No temperature independent crossing point
can be found in those data measured at parallel magnetic fields.
Bottom inset: TCR versus H at temperature of $180~mK$, for
parallel and perpendicular magnetic fields.}
\end{figure}

Figure~\ref{fig:fig3} shows the temperature dependance
$R_{\square}(T)$ for different values of the magnetic field
applied parallel to the thin film plane. At large magnetic fields,
we recover an insulating-like behavior of the sheet resistance, as
in the previous situation. However, a close inspection of the
transition shows that the position of crossing point (R,H), where
a change of sign in TCR occurs, depends on temperature. The color
map of the TCR sign does not show any temperature independent
field scale or kink in the temperature profile of the critical
field. Thus, the transition looks classical, due to the breaking
of Cooper pairs at the temperature-dependant critical field
$H_{c2}$.

An important aspect of the quantum SIT is the supposed existence
of a bosonic insulator -- formed of localized Cooper pairs --
above the critical field. This Bose insulator should only exist
for perpendicular magnetic field configuration; while for parallel
magnetic field, the insulator is of fermion-glass type.
Indications for two different types of insulators is found in the
data.

In perpendicular magnetic field configuration, the inset of
figure~\ref{fig:fig1} shows that TCR has a local negative minimum
that occurs at a field value $H_m\approx 10~kOe$ about two times
the critical field. The disappearance of this minimum at high
temperature (400~mK) indicates that it is related to
superconductivity. The inset figure~\ref{fig:fig3} shows this data
compared with TCR measured in parallel magnetic field. It clearly
appears that the minimum observed in perpendicular magnetic field,
$\approx -200~\Omega K^{-1}$, is well below the TCR measured in
parallel magnetic fields,$\approx -120~\Omega K^{-1}$, and that no
such local minimum exists for the data in parallel magnetic field
configuration. Taken together, these observations suggest that,
for perpendicular magnetic field, superconducting fluctuations are
responsible of the stronger insulating behavior just above the
critical field, which may be the signature of a Bose insulator. If
this interpretation is correct, the subsequent increase of TCR
above $H_m$ is due to the pair-breaking of Cooper pairs; the
bosonic insulator is transformed progressively into a fermionic
insulator toward high magnetic fields. This phenomenon received
much attention recently with the observation of large negative
magnetoresistance in amorphous $In_2O_3$\cite{Steiner05}.

To summarize, we have found that quantum SIT are characterized by
an unambiguous signature -- a kink in the temperature profile of
the critical field. We used this signature to show definitely that
the nature of magnetic field-induced SIT in disordered thin films
of $Nb_{0.15}Si_{0.85}$ depends on the orientation of the magnetic
field with the plane of the film. In perpendicular magnetic
fields, the transition shows the archetypal features of a quantum
SIT : a kink and a plateau in the temperature dependence of the
critical field , a critical scaling behavior and stronger
insulating behavior just above the critical field. In contrast, in
parallel magnetic fields, the SIT looks classical, the
superconductivity disappears at the temperature dependent critical
field $H_{c2}$, due to the breaking of Cooper pairs.

The authors would like to thanks M.
Aprili, A.M. Goldman and S. Vishveshwara for useful discussions.

\bibliography{Super_Insulator}

\end{document}